\documentclass[twocolumn,showpacs,preprintnumbers,amsmath,amssymb,aps,prb]{revtex4}

\usepackage{graphicx}

\begin{document}
\draft
\title{Optical investigations on Y$_{2-x}$Bi$_{x}$Ru$_{2}$O$_{7}$: Electronic
structure evolutions related to the metal-insulator transition}
\author{J. S. Lee, S. J. Moon, and T. W. Noh}
\address{School of Physics and Research Center for Oxide Electronics, Seoul\\
National University, Seoul 151-747, Korea}
\author{T. Takeda}
\address{Division of Materials Science and Engineering, Graduate School of\\
Engineering, Hokkaido University, Hokkaido 060-8628, Japan}
\author{R. Kanno}
\address{Department of Electronic Chemistry, Interdisciplinary Graduate School of Science and Engineering, Tokyo Institute of Technology, Yokohama 226-8502,\\
Japan}
\author{S. Yoshii and M. Sato}
\address{Department of Physics, Division of Material Science, Nagoya University, Furo-cho, Chikusa-ku, Nagoya 464-8602, Japan}
\date{\today }

\begin{abstract}
Optical conductivity spectra of cubic pyrochlore Y$_{2-x}$Bi$_{x}$Ru$_{2}$O$%
_{7}$ (0.0$\leq ${\it x}$\leq $2.0) compounds are investigated. As a
metal-insulator transition (MIT) occurs around {\it x}$=$0.8, large spectral
changes are observed. With increase of {\it x}, the correlation-induced peak
between the lower and the upper Hubbard bands seems to be suppressed, and a
strong mid-infrared feature is observed. In addition, the $p-d$ charge
transfer peak shifts to the lower energies. The spectral changes cannot be
explained by electronic structural evolutions in the simple
bandwidth-controlled MIT picture, but are consistent with those in the
filling-controlled MIT picture. In addition, they are also similar to the
spectral changes of Y$_{2-x}$Ca$_{x}$Ru$_{2}$O$_{7}$ compounds, which is a
typical filling-controlled system. This work suggests that, near the MIT,
the Ru bands could be doped with the easily polarizable Bi cations.
\end{abstract}


\maketitle
\newpage

\section{Introduction}

Pyrochlore ruthenium oxides $A_{2}$Ru$_{2}$O$_{7-\delta }$ [$A$=Bi, Tl, Pb,
Y, and $L$(=Pr-Lu)] are interesting materials, which show numerous
electronic properties depending on the $A$-site ions. While Bi$_{2}$Ru$_{2}$O%
$_{7}$ and Pb$_{2}$Ru$_{2}$O$_{6.5}$ are good metals, Y$_{2}$Ru$_{2}$O$_{7}$
and $\ L_{2}$Ru$_{2}$O$_{7-\delta }$ are insulators.\cite
{Subramanian,Kanno,Yoshii} On the other hand, Tl$_{2}$Ru$_{2}$O$_{7}$ shows
a temperature-dependent metal-insulator transition (MIT) around 125 K.\cite
{Takeda} Therefore, it is clear that the electronic structures, especially
near the Fermi level ($E_{F}$), should have some systematic evolutions
depending on the $A$-site ions. Numerous investigations have been made to
understand how the electronic structures will evolve in these pyrochlore
ruthenates,\cite{Cox,Goodenough,KSLee,Hsu,Ishii,JSLee
PRB,Okamoto,JPark,Kennedy1,Kennedy2} however, there is no real consensus on
this issue yet.

One explanation is based on the fact that the different size of the $A$-site
ions will control structural properties, such as Ru-O-Ru bond angle, which
result in bandwidth changes of the Ru $t_{2g}$ bands. With strong
electron-electron correlation effects,\cite{Cox,Goodenough,KSLee,JSLee
PRB,Okamoto,JPark} these changes could result in a MIT, called the
`bandwidth-controlled MIT'. In the early days, Cox {\it et al}. investigated
Bi$_{2}$Ru$_{2}$O$_{7}$, Y$_{2}$Ru$_{2}$O$_{7}$, and Pb$_{2}$Ru$_{2}$O$%
_{6.5} $ by using photoemission spectroscopy and high-resolution
electron-energy-loss spectroscopy, and found that the density of states at $%
E_{F}$ should decrease in the sequence of Pb$_{2}$Ru$_{2}$O$_{6.5}$, Bi$_{2}$%
Ru$_{2}$O$_{7}$, and Y$_{2}$Ru$_{2}$O$_{7}$.\cite{Cox} They concluded that Y$%
_{2}$Ru$_{2}$O$_{7}$ should be in the insulating state due to
correlation-induced electron localization, namely the Mott insulator. Lee
{\it et al}. calculated the bandwidth of the Ru $t_{2g}$ bands using an
extended Huckel tight binding method, and confirmed that the MIT in the
pyrochlore ruthenates should originate from the change of the relative size
between the correlation energy and bandwidth, which was controlled by the $A$%
-site ion size.\cite{KSLee}

The other explanation is based on the fact that Pb or Bi 6$p$ electrons
should have very large wavefunctions, which could hybridize with the Ru $%
t_{2g}$ wavefunctions. This hybridization could result in large bandwidths
of the Ru $t_{2g}$ bands and a net transfer of charge to the Ru $t_{2g}$
bands,\cite{Hsu,Ishii,JSLee PRB,Okamoto} so the ruthenates could experience
a MIT, called the `filling-controlled MIT'. Hsu {\it et al}. compared the
{\it x}-ray photoemission spectroscopy spectra of Pb$_{2}$Ru$_{2}$O$_{6.5}$
and Bi$_{2}$Ru$_{2}$O$_{7}$ with band-structure calculations and argued that
the unoccupied Pb or Bi 6$p$ states should be closely related to their
metallic conductivity through mixing with the Ru 4$d$ states via the ligand
oxygen 2$p$ states.\cite{Hsu} Later, using the LDA calculation on $A_{2}$Ru$%
_{2}$O$_{7-\delta }$ ($A$=Bi, Tl, and Y), Ishii and Oguchi also obtained
similar results that the Bi 6$p$ and the Tl 6$s$ states could hybridize with
the Ru $t_{2g}$ states and should contribute to the density of states at $%
E_{F}$.\cite{Ishii} Considering the fact that the formal charge valences of
Pb and Bi are +3 just like those of the other pyrochlore ruthenates, the
filling-controlled MIT is rather surprising. Consequently, it is important
to discriminate whether the metallic state of the pyrochlore ruthenates
originates from the large bandwidth of the Ru 4$d$ bands or from the filling
change.

Optical spectroscopy has been used as a powerful tool to investigate
electronic structures of highly correlated electron systems.\cite
{Cooper,KWKim} Several optical studies have been already done on some
pyrochlore ruthenates, and provided some clues to understanding of their
electronic structure changes.\cite{JSLee PRB,Tl2Ru2O7,PhysicaC,Pb2Ru2O65}
Earlier, we compared the optical spectra of Y$_{2}$Ru$_{2}$O$_{7}$, CaRuO$%
_{3}$, SrRuO$_{3}$, and Bi$_{2}$Ru$_{2}$O$_{7}$, and we already showed that Y%
$_{2}$Ru$_{2}$O$_{7}$ should be a Mott insulator and that the metallic
states of the perovskites could be understood in the bandwidth control
picture.\cite{JSLee PRB} However, we stated that the metallic state of the
pyrochlore Bi$_{2}$Ru$_{2}$O$_{7}$ could not be easily understood. [See
comment 32 in Ref. 10.]

In this paper, we report the optical conductivity spectra $\sigma (\omega )$
of Y$_{2-x}$Bi$_{x}$Ru$_{2}$O$_{7}$ (YBRO) ($x$=0.0, 0.5, 1.0, 1.5, and
2.0). YBRO is a good model system to investigate the MIT mechanism of the
pyrochlore ruthenates, since it has both insulating and metallic end
members. In addition, its solid solution can be easily formed in all the
region of $x$, and a MIT occurs around $x$=0.8.\cite{Kanno,Yoshii} When the
YBRO compound becomes metallic, the structural symmetry remains as a cubic;
however, with increase of $x$, the Ru-O bond length decreases and the
Ru-O-Ru bond angle increases.\cite{Kanno} These $x$-dependences are
consistent with those in other pyrochlore ruthenates.\cite
{Kennedy1,Kennedy2,Yamamoto,Kobayashi} From the room temperature $\sigma
(\omega )$ of YBRO, we observed that the variation of $x$ could result in
systematic spectral changes, including large peak shifts and spectral weight
redistributions in a wide energy range up to 5 eV. Using these spectral
changes, we will address the MIT of YBRO from the bandwidth- and
filling-controlled pictures. To clarify our argument further, we will also
compare $\sigma (\omega )$ of YBRO with those of Y$_{2-x}$Ca$_{x}$Ru$_{2}$O$%
_{7}$, which is a typical filling-controlled system.\cite{YCRO}

\section{EXPERIMENTAL}

Polycrystalline Y$_{2-x}$Bi$_{x}$Ru$_{2}$O$_{7}$ samples were synthesized
using the solid state reaction method.\cite{Kanno} Since the pyrochlore
phase has a cubic structure, optical constants of the pyrochlore ruthenates
should be isotropic, so we can determine their optical constants from the
reflectivity measurements of polycrystalline samples. Before reflectivity
measurements, the sample surfaces were polished up to 0.3 $\mu $m. We
measured reflectivity spectra from 5 meV to 30 eV using numerous
spectrophotometers.\cite{JSLee PRB} After the optical measurements, thin
gold films were evaporated on the samples and their reflectivity spectra
were measured again to correct the errors due to scattering from the sample
surfaces.\cite{KK} From the reflectivity spectra, we performed the
Kramers-Kronig (K-K) analysis to obtain $\sigma (\omega )$. In addition, we
also independently measured $\sigma (\omega )$ between 0.7 and 4.0 eV using
spectroscopic ellipsometry. The results of the K-K analysis agreed with the
ellipsometry data, demonstrating the validity of our K-K analysis.\cite{KK}

Sintered polycrystalline samples of Y$_{2-x}$Ca$_{x}$Ru$_{2}$O$_{7}$ were
also synthesized by the solid reaction.\cite{YCRO} The sample densities were
too low for the reflectivity measurements, so we decided to obtain their
absorption spectra by measuring transmittance spectra of Y$_{2-x}$Ca$_{x}$Ru$%
_{2}$O$_{7}$ particles embedded in a KBr matrix, which is transparent up to
around 4 eV. Mixtures of Y$_{2-x}$Ca$_{x}$Ru$_{2}$O$_{7}$ and KBr powders
were mixed together thoroughly and pressed into pellets, whose thicknesses
were about 1 mm. The transmittance spectra were measured between 0.5 and 4.0
eV, and the absorption spectra were evaluated by taking logarithms of the
transmittance spectra and dividing them by the pellet thicknesses.\cite
{Fermi-golden}

\begin{figure}[tbp]
\includegraphics[width=3.3in]{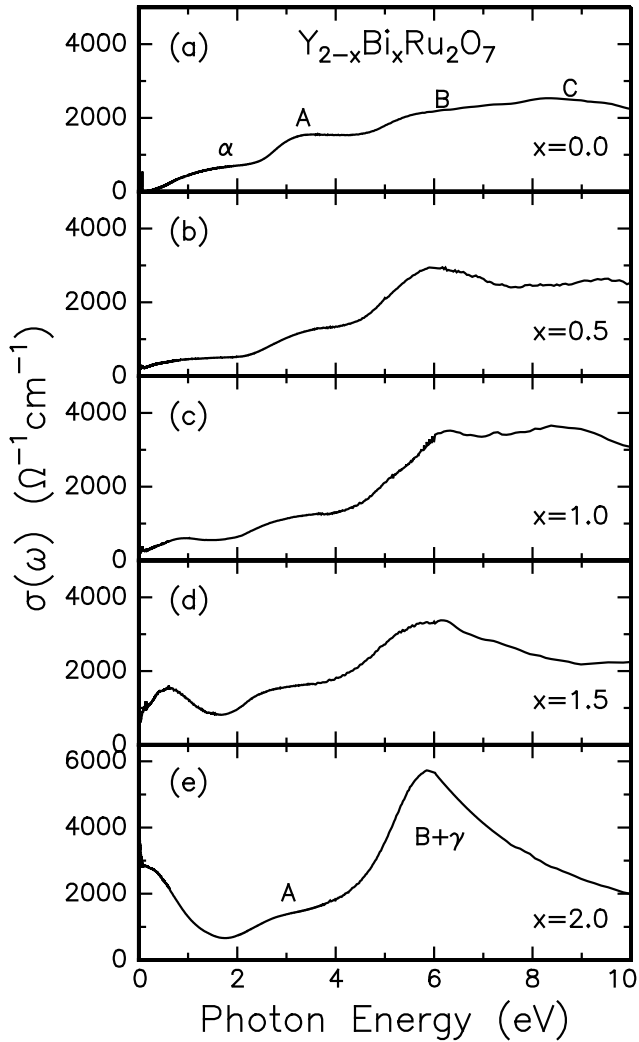}
\caption{Doping-dependent $\protect\sigma (\protect\omega )$ of Y$_{2-x}$Bi$%
_{x}$Ru$_{2}$O$_{7}$ at room temperature up to 10 eV. The indices
of each peak appearing in (a) and (e) indicate the corresponding
optical excitations, displayed in Fig. 2.} \label{Fig:1}
\end{figure}

\section{ASSIGNMENT OF SPECTRAL FEATURES OF Y$_{2-x}$Bi$_{x}$Ru$_{2}$O$_{7}$
BASED ON THE ELECTRONIC STRUCTURES OF END MEMBERS}

Figure 1 shows room temperature $\sigma (\omega )$ of Y$_{2-x}$Bi$_{x}$Ru$%
_{2}$O$_{7}$ ($x$=0.0, 0.5, 1.0, 1.5, and 2.0) up to 10 eV. For $x$=0.0, $%
\sigma (\omega )$ shows an insulating behavior with a small optical gap
around 0.14 eV and the lowest interband transition around 1.6 eV.\cite{Gap}
And, it also shows strong peaks around 3 eV, 6 eV, and 9 eV. As $x$
increases, two intriguing spectral changes can be observed. (1) The spectral
weight below around 1.0 eV increases, and a strong Drude-like peak appears
for $x$=2.0. These behaviors are consistent with the $x$-dependence of
dc-resistivity.\cite{Kanno,Yoshii} (2) While the strength of the 6 eV peak
increases, that of the 9 eV peak decreases and almost disappears for $x$=1.5
and 2.0.
\begin{figure}[tbp]
\includegraphics[width=3.0in]{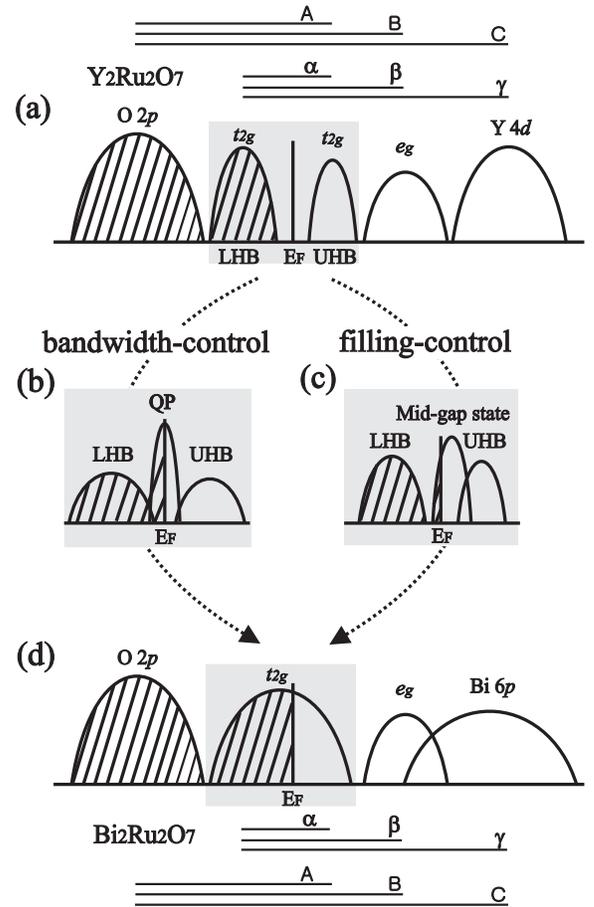}
\caption{Schematic diagrams of the electronic structures of Y$_{2-x}$Bi$_{x}$%
Ru$_{2}$O$_{7}$. (a) corresponds to that for $x$=0, i.e., Y$_{2}$Ru$_{2}$O$%
_{7}$. The occupied and unoccupied Ru $t_{2g}$ states are located near $%
E_{F} $, forming the lower Hubbard band (LHB) and the upper
Hubbard band (UHB), respectively. Outside these Ru $t_{2g}$
states, there are the occupied O 2$p$ state and the unoccupied Ru
$e_{g}$ state and the Y 4$d$ state. (b) sketches possible
electronic structures of the Ru $t_{2g}$ states in the
intermediate region of $x\symbol{126}$1.0, when a metal-insulator
transition occurs through a change of the bandwidth. In the
bandwidth-controlled system, a quasi-particle peak (QP) should
appear between the Hubbard bands. (c) sketches electronic
structures, when a metal-insulator transition occurs due to the
doping. In the filling-controlled system, a mid-gap state should
appear. (d) corresponds to electronic structures for $x$=2.0,
i.e., Bi$_{2}$Ru$_{2}$O$_{7}$. Compared to (a), the Bi 6$p$ state
replaces the Y 4$d$ state, and the $t_{2g}$ states form a single
peak centered near $E_{F}$. In (a) and (d), possible optical
transitions are also indicated as A, B, C, $\protect\alpha $,
$\protect\beta $, and $\protect\gamma $. } \label{Fig:2}
\end{figure}

In order to understand spectral weight changes more easily, let us adopt the
schematic diagrams for the electronic structures of the end members, i.e., Y$%
_{2}$Ru$_{2}$O$_{7}$ and Bi$_{2}$Ru$_{2}$O$_{7}$, which were already
presented in Ref. 10, and also displayed in Figs. 2(a) and (d),
respectively. For Y$_{2}$Ru$_{2}$O$_{7}$, the Ru ion has its formal valence
of 4+ with four $t_{2g}$ electrons. Since it is known as a Mott insulator,
\cite{Cox,JSLee PRB} the partially-filled $t_{2g}$ states split into the
lower Hubbard band (LHB) and the upper Hubbard band (UHB),\cite{RMP 1998} as
shown in the shaded area of Fig. 2(a). The unoccupied $e_{g}$ states are
located above the $t_{2g}$ states by a crystal field splitting energy of
about 3 eV.\cite{JSLee PRB} The occupied O 2$p$ states and the unoccupied Y 4%
$d$ states are located below the LHB and above the $e_{g}$ states,
respectively. On the other hand, Bi$_{2}$Ru$_{2}$O$_{7}$ is close to a band
metallic state,\cite{Cox} where the $t_{2g}$ states form a single
partially-filled band at $E_{F}$, as shown in the shaded area of Fig. 2(d).
Instead of the unoccupied Y 4$d$ states in Y$_{2}$Ru$_{2}$O$_{7}$, Bi$_{2}$Ru%
$_{2}$O$_{7}$ will have the unoccupied Bi 6$p$ states, whose bandwidth might
be much larger due to the extended nature of its corresponding
wavefunctions. Other states, including the O 2$p$ states and the Ru $e_{g}$
states, should remain nearly the same.

In each diagram, possible charge transfer excitations from the O 2$p$ states
are indicated as Transitions A, B, and C, and possible transition from the
Ru $t_{2g}$ states are indicated as Transitions $\alpha $, $\beta $, and $%
\gamma $. For Y$_{2}$Ru$_{2}$O$_{7}$, the lowest excitation around 1.6 eV
can be assigned to the $d-d$ transition between the Hubbard bands, i.e.,
Transition $\alpha $. And, the strong peaks around 3 eV, 6 eV, and 9 eV can
be assigned to Transitions A, B, and C, respectively. For Bi$_{2}$Ru$_{2}$O$%
_{7}$, the coherent peak below 1.5 eV should be attributed to the intraband
transition of the partially-filled $t_{2g}$ band, and the 3 eV peak can be
assigned as Transition A. The position of the 6 eV peak should correspond to
the energies of Transition B and an additional dipole-allowed transition,
i.e., Transition $\gamma $ between the Ru $t_{2g}$ state and the Bi 6$p$
state. The contributions from these two transitions can explain the large
strength of this spectral feature.

The general trends of the high energy spectral changes, shown in Fig. 1, can
be understood based on the electronic structures of Y$_{2}$Ru$_{2}$O$_{7}$
and Bi$_{2}$Ru$_{2}$O$_{7}$. Note that, as the Bi content increases, the 6
eV peak gains its spectral weight, and the 9 eV peak loses its spectral
weight. These systematic $x$-dependences confirm our peak assignments that
the peaks around 6 eV and 9 eV should be related to the Bi ion states and
the Y ion states, respectively. Based on these peak assignments, we can
argue that the spectral features in the low energy region below 5 eV should
come from optical transitions related to the Ru $t_{2g}$ and the O 2$p$
states.

\section{POSSIBLE MODELS FOR ELECTRONIC STRUCTURAL EVOLUTIONS OF Y$_{2-x}$Bi$%
_{x}$Ru$_{2}$O$_{7}$: BANDWIDTH- $vs$. FILLING-CONTROL}

Figures 2(b) and (c) shows the possible electronic structures near $E_{F}$
for the intermediate compounds between Y$_{2}$Ru$_{2}$O$_{7}$ and Bi$_{2}$Ru$%
_{2}$O$_{7}$ in the bandwidth- and the filling-controlled MIT pictures,
respectively. The electronic structures in the high energy regions are not
displayed, since they should exhibit normal $x$-dependences; that is, as $x$
increases, while the O 2$p$ states and the Ru $e_{g}$ states would not be
much affected, the Y 4$d$ states should be replaced by the Bi 6$p$ states.
On the other hand, the details of the Ru $t_{2g}$ states near $E_{F}$ should
change according to the mechanisms of the MIT.

First, let us consider the bandwidth-controlled MIT, where the
quasi-particle (QP) peak should appear near $E_{F}$ between LHB and UHB,\cite
{RMP 1998} as displayed in Fig. 2(b). When the system becomes more metallic,
the QP peak should increase with the reductions of the Hubbard bands. The
corresponding $\sigma (\omega )$ in the low energy region should be composed
of three spectral features: (i) a coherent peak centered at zero energy,
(ii) an incoherent excitation between the Hubbard bands, and (iii) strong $%
p-d$ transitions located at the higher energies. As shown in the inset of
Fig. 3, $\sigma (\omega )$ of CaRuO$_{3}$,\cite{JSLee PRB} which is known to
be a correlated metal, exhibits such spectral features quite well: namely,
the Drude-like peak in the zero energy limit, the correlation-induced peak
around 1.8 eV, and the $p-d$ transition peaks above 3 eV. As the system
approaches the band metallic state, the coherent peak should develop,
accompanied by the reduction of the incoherent excitation.

Second, let us look into the filling-controlled MIT, where a mid-gap state
should appear just below UHB (above LHB) for the hole (electron) doping
case, \cite{RMP 1998} as displayed in Fig. 2(c). Different from the QP peak
in the bandwidth-controlled picture, this mid-gap state is not centered
around $E_{F}$, and it should provide an incoherent transport behavior in a
moderate doping regime due to the disordered-induced carrier localization.
\cite{KWKim} In the hole-doping case, $\sigma (\omega )$ should have two
additional excitations to the mid-gap state from LHB and from the O 2$p$
states, each of which should appear at lower energies than Transitions $%
\alpha $ and A, respectively. As the carrier-doping increases, the
excitations to the mid-gap state should become stronger, accompanied by
spectral weight reductions of Transitions $\alpha $ and A. It should be
noted that the lowest excitation comes from the transition between the LHB
and the mid-gap state, so that it should have a peak center at a finite
energy. Considering large differences in the electronic structures near $%
E_{F}$ and the corresponding $\sigma (\omega )$ between the bandwidth- and
filling-controlled MIT pictures, we will be able to address the possible MIT
mechanism of YBRO by investigating their spectral changes at the low energy
region.

\section{DISCUSSIONS ON THE METAL-INSULATOR TRANSITION MECHANISM OF Y$_{2-x}$%
Bi$_{x}$Ru$_{2}$O$_{7}$}

\subsection{Spectral changes of YBRO below 4.5 eV}

\begin{figure}[tbp]
\includegraphics[width=3.3in]{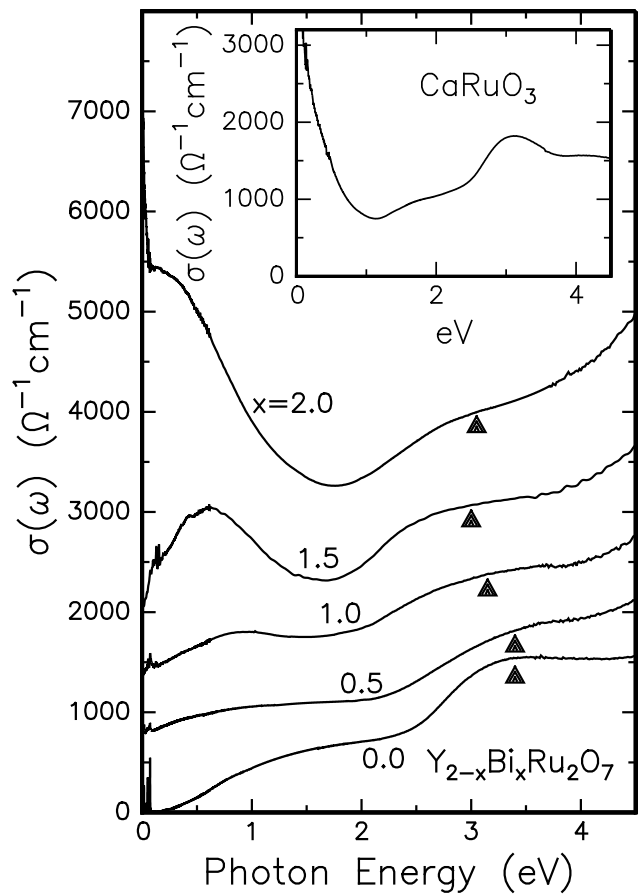}
\caption{Doping-dependent $\protect\sigma (\protect\omega )$ of Y$_{2-x}$Bi$%
_{x}$Ru$_{2}$O$_{7}$ at room temperature up to 4.5 eV. The spectra for $x$%
=0.5, 1.0, 1.5, and 2.0 are shown with an upward shift by 600 $\Omega ^{-1}$%
cm$^{-1}$, 1200 $\Omega ^{-1}$cm$^{-1}$, 1500 $\Omega
^{-1}$cm$^{-1}$, and
2600 $\Omega ^{-1}$cm$^{-1}$, respectively. Inset shows $\protect\sigma (%
\protect\omega )$ of CaRuO$_{3}$ at room temperature.}
\label{Fig:3}
\end{figure}

In order to investigate how the electronic structures of YBRO evolve near $%
E_{F}$ as the MIT occurs, we look into their $\sigma (\omega )$ in the low
energy region. Figure 3 shows $\sigma (\omega )$ of YBRO up to 4.5 eV. As
the MIT occurs, large spectral changes are observed for both the $d-d$ and
the $p-d$ transitions. For the $d-d$ transitions, while $\sigma (\omega )$
of the $x$=0.0 compound exhibits the correlation-induced peak around 1.6 eV,
$\sigma (\omega )$ of the $x$=2.0 compound does not show the
correlation-induced peak but exhibits a strong Drude-like peak extending up
to 1.5 eV. For the $x$=1.0 and 1.5 compounds, strong incoherent mid-infrared
peaks can be observed. On the other hand, the $p-d$ transition shows a
systematic red shift with increase of $x$, as indicated by the solid
triangles. Considering the fact that the spectral features in this energy
region should be mainly contributed by the Ru $t_{2g}$ and the O 2$p$
states, such systematic spectral changes with $x$ indicate that there should
be significant evolutions of these states near the MIT.

\subsection{Discussions based on the bandwidth-control picture}

The evolution of $\sigma (\omega )$ of YBRO in the low energy region, shown
in Fig. 3, cannot be simply explained by the bandwidth-control picture,
which was used by some workers to explain the MIT occurring in the
pyrochlore ruthenates.\cite{Cox,KSLee} According to the diagram in Fig.
2(b), $\sigma (\omega )$ of the intermediate compounds with $x$=1.0 or 1.5
should exhibit a Drude-like peak centered at zero energy and a
correlation-induced $d-d$ transition peak, which is observed for $x$=0.0
around 1.6 eV. Instead of such peak structures, they exhibit just a strong
mid-infrared peak around 1.0 eV for $x$=1.0 (and 0.5 eV for $x$=1.5).
Consequently, the low energy spectral features of the YBRO compounds are
difficult to be understood by those of a correlated metal, located between
the Mott insulator and the band metal.

In addition, the $p-d$ transition peak exhibits an unusual $x$-dependent
evolution, which is different from typical behaviors of the
bandwidth-controlled system. We estimated the energy position $\omega _{p-d}$
of the peak by fitting $\sigma (\omega )$ with a series of the Lorentz
oscillators.\cite{JSLee PRB} As shown in Fig. 4(a), $\omega _{p-d}$ shows a
gradual decrease with increase of $x$. Note that the Ru $t_{2g}$ bandwidth
should be largely influenced by the $p-d$ hybridization between the O 2$p$
and Ru $t_{2g}$ states. If the $p-d$ hybridization plays an important role, $%
\omega _{p-d}$ should be largely determined by the Ru-O bond length $d_{Ru-O}
$: $\omega _{p-d}$ should be inversely proportional to $d_{Ru-O}$.\cite{Cu
Del,Var cuprates,p-d hyb} Namely, the smaller $d_{Ru-O}$ would make the
larger energy splitting between the O 2$p$ states and the Ru $t_{2g}$
states, resulting in larger $\omega _{p-d}$. On the contrary to this
expectation, $\omega _{p-d}$ decreases with decrease of $d_{Ru-O}$,\cite
{Kanno} as shown in Fig. 4(b). This indicates that the observed $d_{Ru-O}$%
-dependence of $\omega _{p-d}$ could not be simply related to the
bandwidth-control effects.

\begin{figure}[tbp]
\includegraphics[width=3.3in]{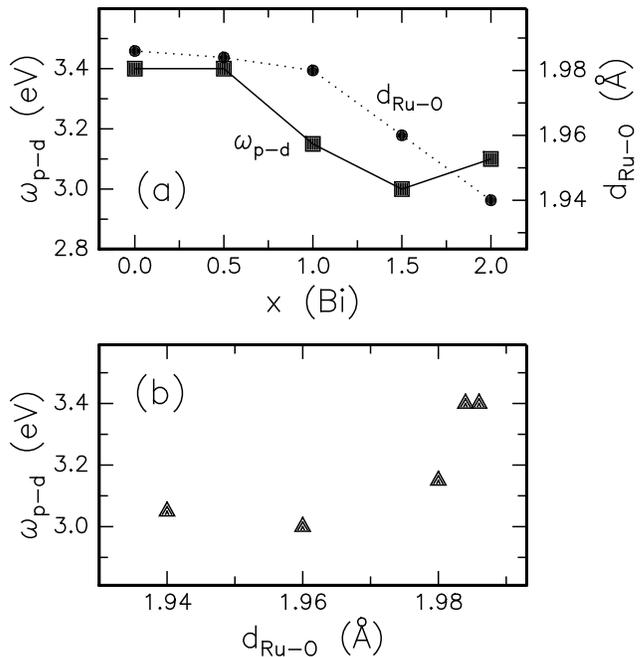}
\caption{(a) Doping-dependent changes of the $p-d$ transition energy $%
\protect\omega _{p-d}$, and (b) Ru-O bond length $d_{Ru-O}$ dependence of $%
\protect\omega _{p-d}$.} \label{Fig:4}
\end{figure}

\subsection{Discussions based on the filling-control picture}

In fact, the filling-controlled MIT picture can provide good explanations to
most of the observed spectral changes of $\sigma (\omega )$ in the low
energy region. As shown in Fig. 3, the $d-d$ transitions and the $p-d$
transitions for the intermediate compounds with $x$=0.5 and 1.0 seem to be
broader than those of end members. And, as $x$ increases, both of them look
like shifting to the lower energies. According to the schematic diagram in
Fig. 2(c), which supposes the hole doping case for the Ru ions, the mid-gap
state becomes stronger with an increase of the hole doping for the Ru ions.
Then the intermediate compounds should have two additional optical
excitations, and each of them should be located just below Transitions $%
\alpha $ and A of the $x$=0.0 compound. This explains why the $d-d$ and the $%
p-d$ transitions for the intermediate compounds become broader. If the
finite widths of the peaks are taken into consideration, the apparent red
shifts of the $d-d$ and the $p-d$ transitions could be attributed to the
increased spectral weights of such additional excitations.

Moreover, the $d_{Ru-O}$-dependence of $\omega _{p-d}$, shown in Fig. 4(b),
are also consistent in the filling-controlled picture. As $x$ increases, the
amount of hole doping for the Ru ions increases, so $d_{Ru-O}$ could
decrease.\cite{Kennedy1,Kennedy2} In addition, with the increase of $x$, the
mid-gap state becomes stronger resulting in the apparent redshifts of $%
\omega _{p-d}$, as we addressed in the above paragraph. Therefore, $\omega
_{p-d}$ will decrease at smaller $d_{Ru-O}$.

Although the filling-controlled MIT picture is consistent with the observed
spectral changes of YBRO, it should be noted that the low frequency spectral
distribution of YBRO is too broad to clearly identify the appearance of the
mid-gap state. In recent photoemission studies, Park {\it et al}. observed
an increase of the spectral weight near $E_{F}$,\cite{JPark} but the
spectral features in the photoemission spectra were also too broad to
clearly distinguish the appearance of the mid-gap state. The broad spectral
features in the mid-infrared peak and the photoemission spectra might be
related to the extended nature of the Bi 6$p$ orbitals,\cite{Hsu,Ishii,JSLee
PRB} which are hybridized with the Ru $t_{2g}$ bands.

\subsection{Comparison with the spectral changes of Y$_{2-x}$Ca$_{x}$Ru$_{2}$%
O$_{7}$, a filling-controlled system}

To obtain further supports for the possible scenario of the YBRO MIT
mechanism, we investigated absorption spectra of Y$_{2-x}$Ca$_{x}$Ru$_{2}$O$%
_{7}$ (YCRO), which shows a MIT around $x$=0.5.\cite{YCRO} Since Y and Ca
ions have different ionic states, as 3+ and 2+, respectively, the
hole-doping for the Ru 4$d$ states should occur and increase with the
increase of $x$. Therefore, the YCRO compounds will work as a model system
with a clear filling-controlled MIT.

\begin{figure}[tbp]
\includegraphics[width=3.3in]{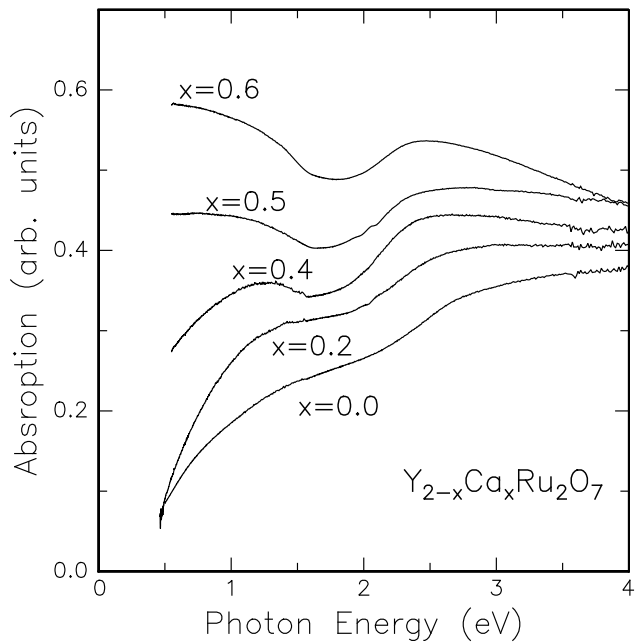}
\caption{Absorption spectra of Y$_{2-x}$Ca$_{x}$Ru$_{2}$O$_{7}$ at
room temperature. All of the spectra are shown in arbitrary
units.\ } \label{Fig:5}
\end{figure}

Figure 5 shows the absorption spectra of the YCRO compounds
between 0.5 and 4 eV. The spectra below 0.5 eV could not be
measured due to the multi-phonon absorption of the KBr pellet. For
the Y$_{2}$Ru$_{2}$O$_{7}$ sample, its absorption spectrum is
quite similar to its $\sigma (\omega )$, displayed in the bottom
of Fig. 3, demonstrating the validity of our approaches using
transmittance spectra. It is quite interesting to find that
$x$-dependent changes of the absorption spectra for YCRO compounds
are quite similar with those of $\sigma (\omega )$ for the YBRO
compounds, shown in Fig. 3. Characteristic spectral features in
YBRO, i.e., the red shifts of the $d-d$ transition peak and the
$p-d$ transition peak, can also be observed in the absorption
spectra of YCRO. These similarities of optical spectra between
YBRO and YCRO strongly support our proposal that the MIT of the
YBRO should originate from the doping effects by the Bi cations.

\subsection{Detailed mechanism of the doping effects}

As $x$ increases, the Ru $t_{2g}$ bandwidth should become larger through the
hybridization between the Bi 6$p$ states and the Ru $t_{2g}$ states.\cite
{Hsu,Ishii,JSLee PRB,HDKim} However, our optical investigations clearly
demonstrated that the hole doping will play a more important role in
determining the MIT in the YBRO compounds. How will the Bi substitution
result in the hole doping? According to the LDA calculations on the
pyrochlore ruthenates by Ishii and Oguchi,\cite{Ishii} the electron orbitals
at the Y site will not be mixed with the Ru $t_{2g}$ orbitals, but the
unoccupied Bi 6$p$ orbitals should be strongly hybridized with the Ru $%
t_{2g} $ orbitals. These results imply that the substitution of the Y ion
with the Bi ion could give rise to the hole-doping into the Ru $t_{2g}$
bands. In particular, the extended wavefunction of 6$p$ states of the easily
polarizable Bi cation makes the self-doping occur.

According to the LDA calculations by Ishii and Oguchi,\cite{Ishii} an
antibonding band of the Tl 6$s$ and the O 2$p$ orbitals in pyrochlore Tl$%
_{2} $Ru$_{2}$O$_{7}$ is expected to cross $E_{F}$, which could provide a
self-doping to the Ru $t_{2g}$ states. In our previous optical studies on Tl$%
_{2}$Ru$_{2}$O$_{7}$,\cite{Tl2Ru2O7,PhysicaC} we reported an incoherent
mid-infrared peak exhibiting unusual temperature-dependences. Contrary to
the YBRO case, the mid-infrared peak feature could easily be distinguishable
from other peaks, and its temperature-dependences are similar to the
doping-dependences of the mid-infrared peak in an externally doped Mott
insulator. \cite{Kasuya:V,Taguchi:Ti} Therefore, we explained the
temperature-dependent MIT of Tl$_{2}$Ru$_{2}$O$_{7}$ in terms of a
self-doping effect for the Ru ions by the easily polarizable Tl cation.\cite
{Tl2Ru2O7,PhysicaC} Note that similar self-doping has been observed for Tl-
ad Bi-based high-temperature superconductors.\cite
{Doping1,Doping2,Doping3,Doping4}

It should be also noted that the geometrical frustration of the pyrochlore
compounds might play important roles in determining their electronic
structures.\cite{Frustration1,Frustration2,Frustration3} In the pyrochlore
structures where the magnetic ions are located at vertices of tetrahedra,
the antiferromagnetic interaction will experience a strong geometrical
frustration. When the frustration is very strong, heavy quasiparticles are
formed around the Fermi level by suppressing the short-range
antiferromagnetic fluctuations. By decreasing the frustration, the heavy
quasiparticle band splits, and the pseudogap begins to develop around the
Fermi level. We expect that such geometrical frustration effects could also
influence the spectral evolutions of YBRO. However, this goes beyond our
current investigation. Further studies are strongly desirable to understand
how the geometrical frustration should modify the electronic structure
changes in the bandwidth-control and the filling-control pictures.

\section{SUMMARY}

We reported the doping-dependent optical conductivity spectra of the cubic
pyrochlore Y$_{2-x}$Bi$_{x}$Ru$_{2}$O$_{7}$ (0.0$\leq ${\it x}$\leq $2.0),
and investigated the mechanism of a metal-insulator transition of these
alloy compounds. We reported that the spectral features below 5 eV exhibit
large changes with a variation of $x$. We demonstrated that they should be
understood in terms of the change in the filling state of the Ru ions.
Together with the previous optical studies on Tl$_{2}$Ru$_{2}$O$_{7}$,\cite
{Tl2Ru2O7,PhysicaC} this work suggests that the metallic behaviors of some
pyrochlore ruthenates, such as Bi$_{2}$Ru$_{2}$O$_{7}$ and Tl$_{2}$Ru$_{2}$O$%
_{7}$ (above 125 K), should be understood in terms of the self-doping for
the Ru 4$d$ states by the easily polarizable cations, i.e., Bi and Tl.

\acknowledgments
We would like to thank S. Fujimoto for valuable discussions. This work was
supported by the Ministry of Science and Technology through the Creative
Research Initiative program, and by KOSEF through the Center for Strongly
Correlated Materials Research. The experiments at PLS were supported by MOST
and POSCO.


\begin{references}
\bibitem{Subramanian}  M. A. Subramanian, G. Aravamudan, and G. V. Subba
Rao, Pro. Solid State Chem. {\bf 15}, 55 (1983).

\bibitem{Kanno}  R. Kanno, Y. Takeda, T. Yamamoto, Y. Kawamoto, and O.
Yamamoto, J. Solid State Chem. {\bf 102}, 106 (1993).

\bibitem{Yoshii}  S. Yoshii and M. Sato, J. Phys. Soc. Jpn. {\bf 68}, 3034
(1999).

\bibitem{Takeda}  T. Takeda, M. Nagata, H. Kobayaxhi, R. Kanno, Y. Kawamoto,
M. Takano, T. Kamiyama, F. Izumi, and A. W. Sleight, J. Solid State Chem.
{\bf 140}, 182 (1998).

\bibitem{Cox}  P. A. Cox, R. G. Egdell, J. B. Goodenough, A. Hamnett, and C.
C. Naish, J. Phys. C {\bf 16}, 6221 (1983).

\bibitem{Goodenough}  J. B. Goodenough, A. Hamnett, and D. Telles, in
Localization and M-I Transition, edited by H. Fritzshe and D. Adler (Plenum,
New York, 1985), p. 161.

\bibitem{KSLee}  K.-S. Lee, D.-K. Seo, and M.-H. Whangbo, J. Solid State
Chem. {\bf 131}, 405 (1997).

\bibitem{Hsu}  W. Y. Hsu, R. V. Kasowski, T. Miller, and T.-C. Chiang, Appl.
Phys. Lett. {\bf 52}, 792 (1988).

\bibitem{Ishii}  F. Ishii and T. Oguchi, J. Phys. Soc. Jpn. {\bf 69}, 526
(2000).

\bibitem{JSLee PRB}  J. S. Lee, Y. S. Lee, T. W. Noh, K. Char, Jonghyurk
Park, S.-J. Oh, J.-H. Park, C. B. Eom, T. Takeda, and R. Kanno, Phys. Rev. B
{\bf 64}, 245107 (2001).

\bibitem{Okamoto}  J. Okamoto, T. Mizokawa, A. Fujimori, T. Takeda, R.
Kanno, F. Ishii, and T. Oguchi, Phys. Rev. B {\bf 69}, 035115 (2004).

\bibitem{JPark}  J. Park, K. H. Kim, H.-J. Noh, S.-J. Oh, J.-H. Park, H.-J.
Lin, and C.-T. Chen, Phys. Rev. B {\bf 69}, 165120 (2004).

\bibitem{Kennedy1}  B. J. Kennedy and T. Vogt, J. Solid State Chem. {\bf 126}%
, 261 (1996).

\bibitem{Kennedy2}  B. J. Kennedy, Physica B {\bf 241-243}, 303 (1998).

\bibitem{Cooper}  S. L. Cooper, Struct. Bonding (Berlin) {\bf 98}, 164
(2001), and references therein.

\bibitem{KWKim}  K. W. Kim, J. S. Lee, T. W. Noh, S. R. Lee, and K. Char,
Phys. Rev. B {\it in press}.

\bibitem{Tl2Ru2O7}  J. S. Lee, Y. S. Lee, K. W. Kim, T. W. Noh, J. Yu, T.
Takeda, and R. Kanno, Phys. Rev. B {\bf 64}, 165108 (2001).

\bibitem{PhysicaC}  J. S. Lee, Y. S. Lee, K. W. Kim, T. W. Noh, J. Yu, Y.
Takeda, and R. Kannoo, Physica C {\bf 364-365}, 632 (2001).

\bibitem{Pb2Ru2O65}  P. Zheng, N. L. Wang, J. L. Luo, R. Jin, and D.
Mandrus, Phys. Rev. B {\bf 69}, 193102 (2004).

\bibitem{Yamamoto}  T. Yamamoto, R. Kanno, Y. Takeda, O. Yamamoto, Y.
Kawamoto, and M. Takano, J. Solid State Chem. {\bf 109}, 372 (1994).

\bibitem{Kobayashi}  H. Kobayashi, R. Kanno, Y. Kawamoto, T. Kamiyama, F.
Izumi, and A.W. Sleight, J. Solid State Chem. {\bf 114}, 15 (1995).

\bibitem{YCRO}  S. Yoshii, K. Murata, and M. Sato, Journal of Physics and
Chemistry of Solids {\bf 62}, 129 (2001).

\bibitem{KK}  H. J. Lee, J. H. Jung, Y. S. Lee, J. S. Ahn, T. W. Noh, K. H.
Kim, and S.-W. Cheong, Phys. Rev. B. {\bf 60}, 5251 (1999).

\bibitem{Fermi-golden}  F. Wooten, {\it Optical Properties of Solids}
(Academic press, New York and Londen, 1972).

\bibitem{Gap}  J. S. Lee, T. W. Noh, J. S. Bae, I.-S. Yang, T. Takeda, and
R. Kanno, Phys. Rev. B {\bf 69}, 214428 (2004).

\bibitem{RMP 1998}  M. Imada, A. Fujimori, and Y. Tokura, Rev. Mod. Phys.
{\bf 70}, 1039 (1998), and references therein.

\bibitem{Cu Del}  S. L. Cooper, G. A. Thomas, A. J. Millis, P. E. Sulewski,
J. Orenstein, D. H. Rapkine, S-W. Cheong, and P. L. Trevor, Phys. Rev. B
{\bf 42}, 10 785 (1990).

\bibitem{Var cuprates}  Y. Ohta, T. Tohyama, and S. Maekawa, Phys. Rev.
Lett. {\bf 66}, 1228 (1991), and references therein.

\bibitem{p-d hyb}  J. S. Lee, Y. S. Lee, T. W. Noh, S. Nakatsuji, H.
Fukazawa, R. S. Perry, Y. Maeno, Y. Yoshida, S. I. Ikeda, J. Yu, and C. B.
Eom, Phys. Rev. B {\bf 70}, 085103 (2004).

\bibitem{HDKim}  Hyeong-Do Kim, Han-Jin Noh, K. H. Kim, and S.-J. Oh, Phys.
Rev. Lett. {\bf 93}, 126404 (2004).

\bibitem{Kasuya:V}  M. Kasuya, Y. Tokura, and T. Arima, Phys. Rev. B {\bf 47}%
, 6197 (1993).

\bibitem{Taguchi:Ti}  Y. Taguchi, Y. Tokura, T. Arima, and F. Inaba, Phys.
Rev. B {\bf 48}, 511 (1993).

\bibitem{Doping1}  H. Krakauer and W. E. Pickett, Phys. Rev. Lett. 60, 1665
(1988).

\bibitem{Doping2}  D. R. Hamann, and L. F. Mattheiss, Phys. Rev. B 38, 5012
(1988).

\bibitem{Doping3}  S. Massidda, J. Yu, and A. J. Freeman, Physica C 152, 251
(1988); J. Yu, S. Massidda, and A. J. Freeman, Physica C 152, 273 (1988).

\bibitem{Doping4}  R. Retoux, F. Studer, C. Michel, B. Raveau, A. Fontaine,
and E. Dartyge, Phys. Rev. B 41, 193 (1990).

\bibitem{Frustration1}  Y. Imai and N. Kawakami, Phys. Rev. B {\bf 65},
233103 (2002).

\bibitem{Frustration2}  S. Fujimoto, Phys. Rev. B {\bf 64}, 85102 (2001); S.
Fujimoto, Phys. Rev. Lett. {\bf 89}, 226402 (2002).

\bibitem{Frustration3}  J. Hopkinson and P. Coleman, Phys. Rev. Lett. {\bf 89%
}, 267201 (2002).
\end{references}
\end{document}